\documentclass{aip-cp}

\usepackage[numbers]{natbib}
\usepackage{rotating}
\usepackage{graphicx}

\begin{document}

\title{Dynamics of quasifission}

\author[aff1]{A.S. Umar\corref{cor1}}
\author[aff1]{V.E. Oberacker}
\author[aff2]{C. Simenel}

\affil[aff1]{Department of Physics and Astronomy, Vanderbilt University, Nashville, Tennessee 37235, USA}
\affil[aff2]{Department of Nuclear Physics, RSPE, Australian National University, Canberra, ACT 0200, Australia}
\corresp[cor1]{Corresponding author: umar@compsci.cas.vanderbilt.edu}

\maketitle

\begin{abstract}
Quasifission is the primary reaction mechanism that limits the formation of
superheavy nuclei and consequently an important ingredient for choosing the best
target-projectile combinations for the heavy element searches.
Quasifission is characterized by nuclear contact-times that are
much longer than those found in deep-inelastic reactions,
resulting in a substantial mass and charge transfer. In this manuscript we employ the fully microscopic time-dependent Hartree-Fock (TDHF) theory
to study quasifission. New results are presented for the $^{48}$Ca+$^{249}$Bk system.

\end{abstract}

\section{INTRODUCTION}

The search for new elements is one of the most challenging research
areas of nuclear physics~\cite{hofmann2000,oganessian2007}.
The discovery of a region of the nuclear chart that can sustain the so
called \textit{superheavy elements} (SHE) has lead to intense experimental activity
resulting in the discovery of elements with atomic numbers as large as $Z=118$.
The experiments to discover these new elements are notoriously difficult, with
production cross-sections in pico-barns.
Of primary importance for the experimental investigations appear to be the choice
of target-projectile combinations that have the highest probability for forming
a compound nucleus that results in the production of the desired element.
Therefore it is important to understand the details of the reaction dynamics of these systems.
For heavy systems leading to superheavy
formations, the evaporation residue cross-section is dramatically reduced due to the
quasi-fission (QF) and fusion-fission
processes thus making the capture cross-section to
be essentially the sum of these two cross-sections, with QF occurring at a shorter time-scale but still  much longer
than those found in deep-inelastic reactions and with a very substantial
mass and charge transfer~\cite{toke1985}.

Within the last few years the time-dependent Hartree-Fock (TDHF) approach~\cite{negele1982,simenel2012} has been 
utilized for studying the dynamics of
quasifission~\cite{golabek2009,kedziora2010,wakhle2014,oberacker2014,kalee2015}
and scission dynamics~\cite{simenel2014a,scamps2015,goddard2015}. Particularly, the study of quasifission is showing a great promise to provide
insight based on very favorable comparisons with experimental data. 
Similarly, an extension of TDHF called the density-constrained TDHF~\cite{umar2006b}
(DC-TDHF) have
been used to obtain microscopic potential barriers and capture cross-sections for
superheavy~\cite{umar2010a} and lighter systems~\cite{oberacker2010,keser2012,simenel2013a}.
In this article we will
focus on the TDHF studies of quasifission observables and related quantities.

\section{RESULTS}
During the past several years it has become feasible to perform TDHF calculations
on a three-dimensional (3D) Cartesian grid with no symmetry restrictions
and with much more accurate numerical methods~\cite{bottcher1989,maruhn2014}.
In the present TDHF calculations we use the Skyrme SLy4d
energy density functional (EDF)~\cite{kim1997}
including all of the relevant time-odd terms in the mean-field Hamiltonian.
First we generate very accurate static HF wave functions for the two nuclei on the
3D grid.
The initial separation of the two nuclei is $30$~fm. In the second
step, we apply a boost operator to the single-particle wave functions. The time-propagation
is carried out using a Taylor series expansion (up to orders $10-12$) of the
unitary mean-field propagator, with a time step $\Delta t = 0.4$~fm/c.
By virtue of long contact-times for quasi-fission and the energy
and impact parameter dependence these calculations require extremely long CPU times.
\begin{figure}[!htb]
	\centerline{\includegraphics*[scale=0.4]{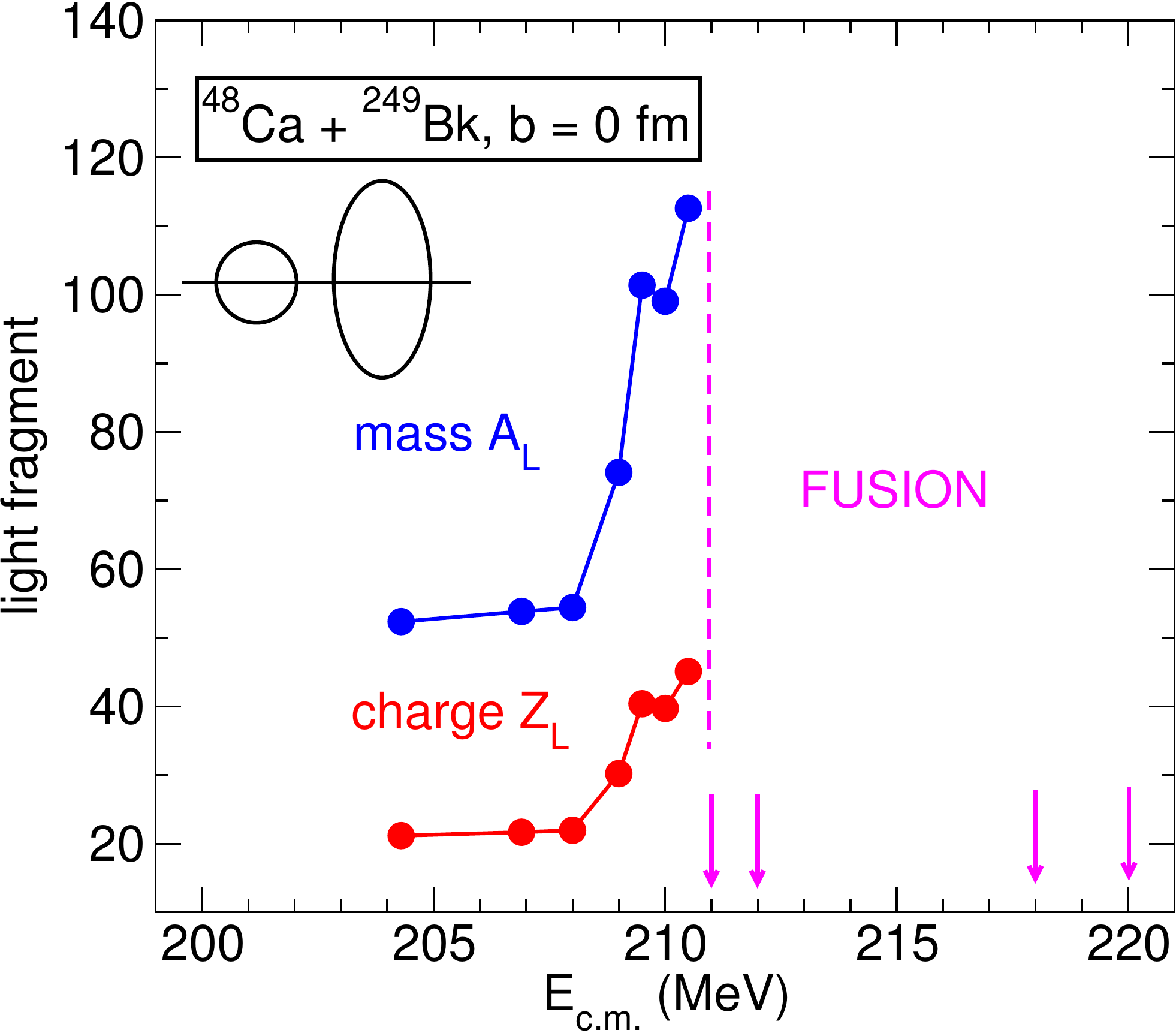}}
	\caption{\protect Mass and charge of the light fragment as a function of $E_{c.m.}$
		for central collisions of $^{48}$Ca with the side of $^{249}$Bk. The
		purple arrows indicate the energies where fusion has been found in TDHF.}
	\label{fig:fig0}
\end{figure}

Figure~\ref{fig:fig0} shows the mass and charge of the light fragment as a function of $E_{c.m.}$
for central collisions of $^{48}$Ca with the side of $^{249}$Bk. For energies below
$E_{c.m.}=204$~MeV we get quasielastic collisions whereas for energies above 211~MeV we see
fusion, which we define arbitrarily as reactions with contact times exceeding 35~zs. Naturally, non-central impact
parameters can show quasifission in the range where we see fusion.
What we observe is that for central collisions quasifission is limited to a small range
of energies $E_{c.m.}=209-211$~MeV.
\begin{figure}[!htb]
	\centerline{
		\includegraphics*[scale=0.4]{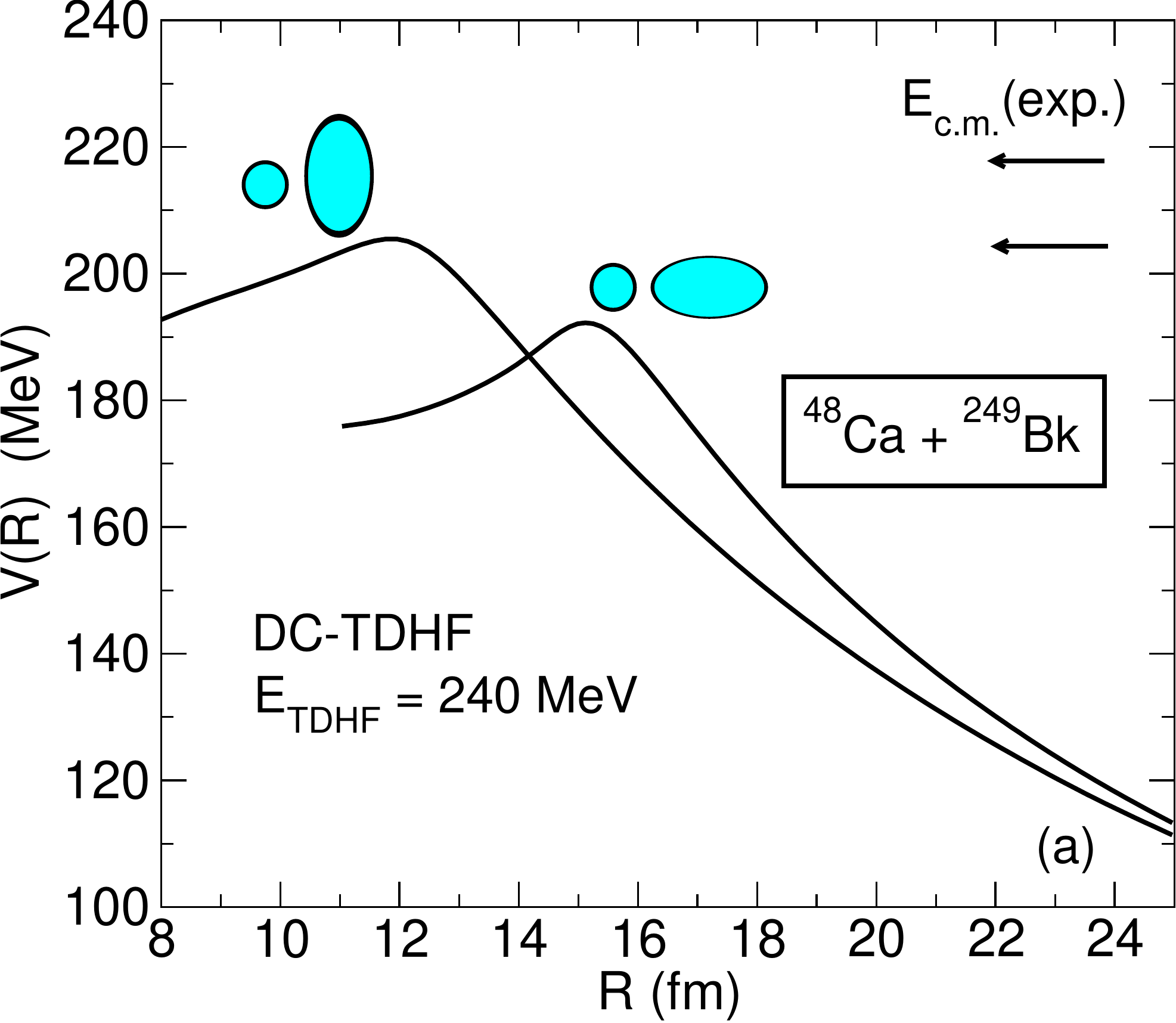}\hspace{0.2in}
		\includegraphics*[scale=0.4]{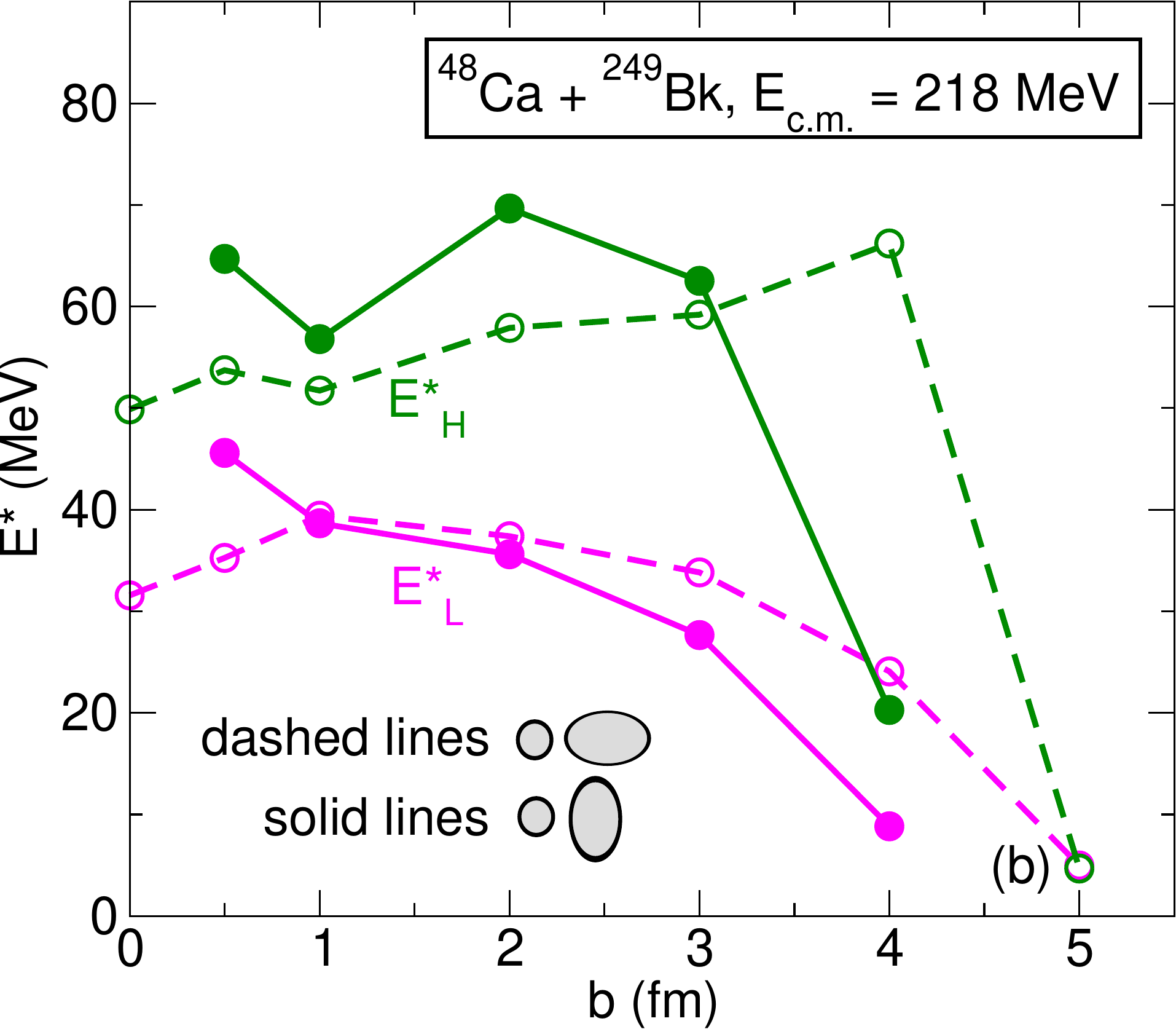}}
	\caption{\protect (a) Nucleus-nucleus potential, $V(R)$,  for the $^{48}$Ca+$^{249}$Bk system
		obtained from DC-TDHF calculation for selected orientation angles of the $^{249}$Bk nucleus. Also shown are the experimental c.m. energies.
		(b) Excitation energy, $E^{*}$, of the heavy and light fragments as a function
		of impact parameter and for two orientations of the $^{249}$Bk nucleus
		calculated at $E_{\mathrm{c.m.}}= 218$~MeV.}
	\label{fig:fig1}
\end{figure}

In Fig.~\ref{fig:fig1}a we plot the microscopic DC-TDHF potential barriers obtained
for the $^{48}$Ca+$^{249}$Bk system. The two barriers depict the two extreme orientations
of the $^{249}$Bk nucleus. Also, shown are the experimental energies at which this reaction has been studied~\cite{khuyagbaatar2014}.
As expected the polar or tip orientation of $^{249}$Bk results in a significantly lower
barrier. The higher experimental energy is
above both barriers but the lower experimental energy is below the barrier for the
equatorial or side orientation of $^{249}$Bk.
Recently, we have developed an extension to TDHF theory via the use
of a density constraint to calculate  excitation energy of {\it each fragment} directly from the
TDHF density evolution.
This gives us new information on the repartition of the excitation energy between the heavy and light fragments
which is not available in standard TDHF calculations.
In Fig.~\ref{fig:fig1}b we show the excitation energies of the
heavy and light fragments which contain approximately 60~MeV and 40~MeV of excitation energy,
respectively, for impact parameters corresponding to quasifission but drop rapidly for
deep-inelastic reactions at larger impact parameters.

\begin{figure}[!htb]
	    \centerline{
		\includegraphics*[scale=0.4]{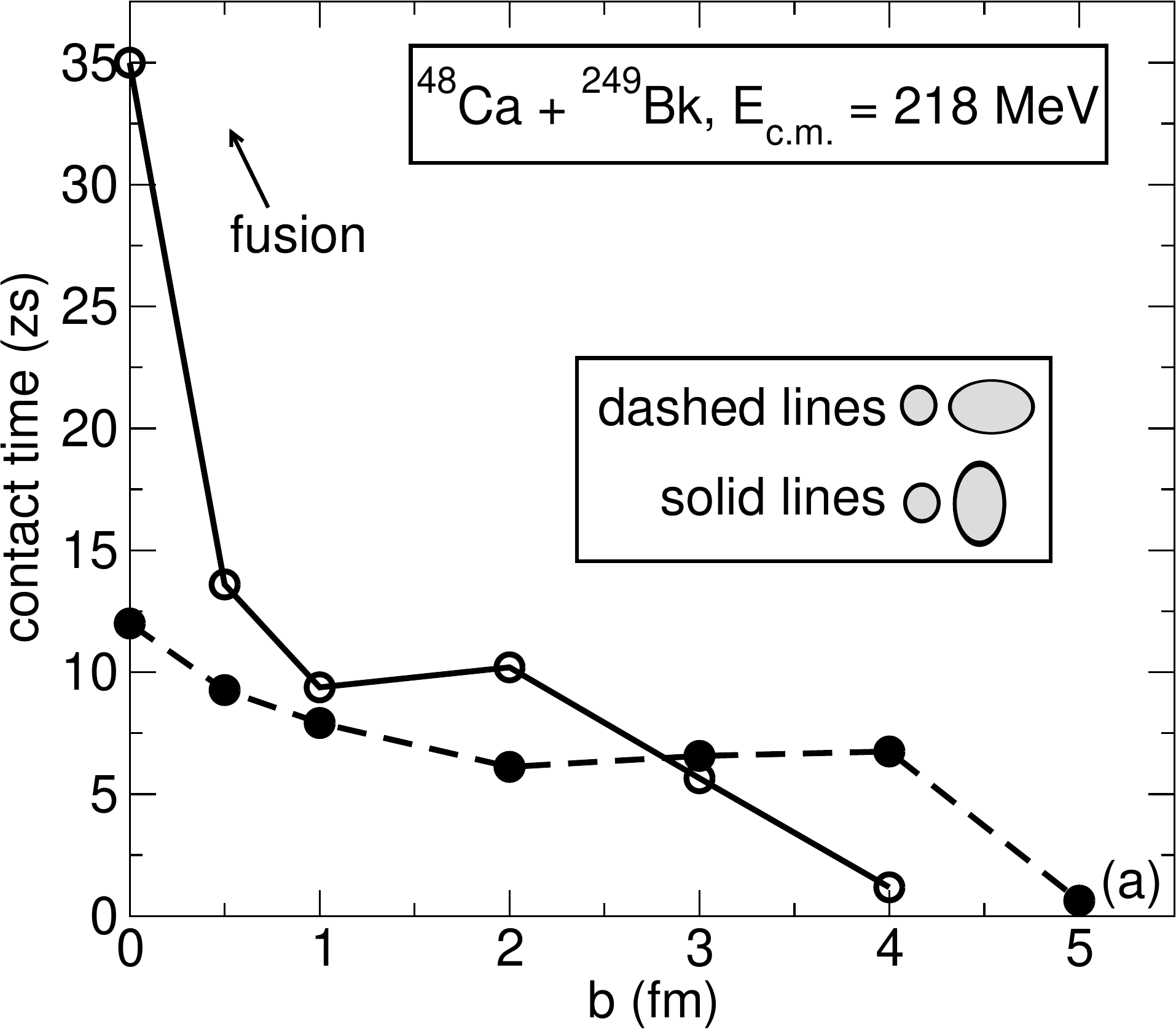}\hspace{0.2in}
		\includegraphics*[scale=0.4]{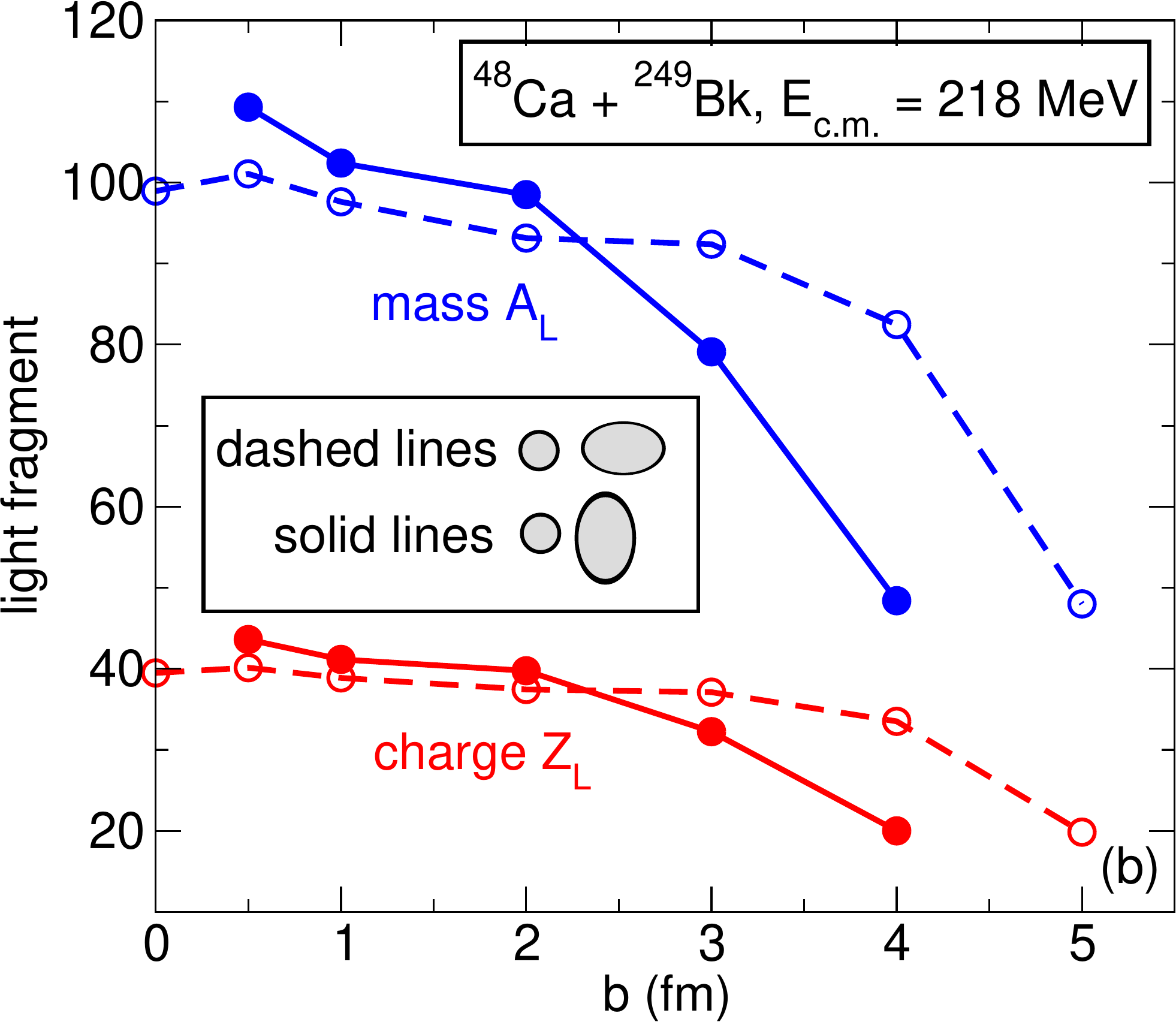}}
	\caption{\protect (a) Contact time and (b) mass and
		charge of the light fragment for the $^{48}$Ca+$^{249}$Bk system as a function of impact parameter for the two
		orientations of the $^{249}$Bk nucleus calculated at the c.m. energy of $E_{\mathrm{c.m.}}= 218$~MeV.}
	\label{fig:fig2}
\end{figure}

Figure~\ref{fig:fig2}a shows the contact time as a function of impact parameter at the
same c.m. energy. We see that the contact times are $8-30$~zs for impact parameters
resulting in quasifission and falls sharply for fragments produced in deep-inelastic
collisions. Figure~\ref{fig:fig2}b show the mass and charge of the light fragment for the $^{48}$Ca+$^{249}$Bk system as a function of impact parameter for the two
orientations of the $^{249}$Bk nucleus calculated at the c.m. energy of $E_{\mathrm{c.m.}}= 218$~MeV. As expected quasifission is identified with large mass and charge transfer,
in this case corresponding to the doubling of the charge from 20 to 40 and mass from 48
to 100.
It is also interesting to note the atypical value of the contact time at impact parameter
$b=2$~fm in Fig.~\ref{fig:fig2}a in comparison to the neighboring impact parameters.
Fig.~\ref{fig:fig2}b shows that in this region
the light fragment is a neutron rich Zr isotope with $A\approx 102-106$.
The microscopic evolution of the shell structure seems to have a tendency to
form a composite with a longer lifetime when the light fragment is in this region.
This was also discussed for the case of $^{40,48}$Ca$+^{238}$U quasifission study
of Ref.~\cite{oberacker2014}. In Ref.~\cite{oberacker2014} this was explained as
being due to the presence of strongly bound deformed isotopes of Zr in this
region~\cite{oberacker2003,blazkiewicz2005}.

\section{SUMMARY}
We have shown recent quasifission results for the $^{48}$Ca+$^{249}$Bk system. Further
calculations are underway to obtain a full range of observables including mass-angle
distributions and fragment TKE's.
Recent TDHF calculations of phenomena related to superheavy element searches show that TDHF can be
a valuable tool for elucidating some of the underlying physics for these reactions.
As a microscopic theory with no free parameters, where the effective nucleon-nucleon
interaction is only fitted to the static properties of a few nuclei, these results
are very promising.

\section{ACKNOWLEDGMENTS}
This work has been supported by the U.S. Department of Energy under grant No.
DE-FG02-96ER40975 and DE-SC0013847 with Vanderbilt University and by the
Australian Research Council Grant No. FT120100760.

\bibliographystyle{aipnum-cp}
\bibliography{AIP_proc.bib}

\end{document}